\PassOptionsToPackage{numbers,sort&compress}{natbib} 
\documentclass[%
preprint,
titlepage,
superscriptaddress,
nofootinbib,
 amsmath,amssymb,
 aps,
]{revtex4-2}

\usepackage[dvipdfmx]{graphicx}
\usepackage{subcaption}
\usepackage{dcolumn}
\usepackage{bm}
\usepackage{threeparttable}
\usepackage{caption}
\usepackage[colorlinks=true, citecolor=blue, linkcolor=blue, urlcolor=blue]{hyperref}
\begin{document}

\preprint{}

\title{
Simple Analytical Solutions of the Wheeler-DeWitt Equation in the Classical Hamilton-Jacobi Limit
}

\author{Naoto Maki}
\affiliation{Department of Astronomical Science, The Graduate University for Advanced Studies (SOKENDAI), 2-21-1 Osawa, Mitaka, Tokyo 181-8588, Japan}
\affiliation{Division of Science, National Astronomical Observatory of Japan, 2-21-1 Osawa, Mitaka, Tokyo 181-8588, Japan}
\author{Chia-Min Lin}
\affiliation{Fundamental General Education Center, National Chin-Yi University of Technology, Taichung 411030, Taiwan, R.O.C.}

\author{Kazunori Kohri}
\affiliation{Division of Science, National Astronomical Observatory of Japan, 2-21-1 Osawa, Mitaka, Tokyo 181-8588, Japan}
\affiliation{Department of Astronomy, The University of Tokyo, Bunkyo-ku, Hongo, Tokyo 113-0033, Japan}
\affiliation{Department of Astronomical Science, The Graduate University for Advanced Studies (SOKENDAI), 2-21-1 Osawa, Mitaka, Tokyo 181-8588, Japan}
\affiliation{Theory Center, IPNS, KEK,
1-1 Oho, Tsukuba, Ibaraki 305-0801, Japan}
\affiliation{Kavli IPMU (WPI), UTIAS, The University of Tokyo, Kashiwa, Chiba 277-8583, Japan}

\date{\today}

\begin{abstract}
We investigate the Wheeler-DeWitt equation for a flat, homogeneous, and isotropic Universe containing a canonical scalar field with a potential. We show that under the constraint $|\Psi|=1$, where the Wheeler-DeWitt equation exactly becomes the classical Hamilton-Jacobi equation, the form of the potential is completely determined depending on the value of the operator ordering parameter. Furthermore, we demonstrate that the classified potentials admit simple forms, such as the exponential, quadratic with a negative cosmological constant, and cosine-type potential with a negative cosmological constant. Several of these have already been explored in the context of inflation or dark energy. Finally, focusing on the system with the cosine-type potential and a negative cosmological constant in the classified potentials, we derive the analytical solutions for the scale factor and the scalar field and discuss the cosmological implications.
\end{abstract}

\maketitle

\section{Introduction}
\label{sec:Introduction}
Quantum cosmology is the framework that aims to formulate a quantum theory of gravity by describing the Universe with a wave function $\Psi$ \cite{DeWitt:1967yk,Hartle:1983ai,Vilenkin:1984wp}. The wave function of the Universe, $\Psi$, is governed by the Wheeler-DeWitt equation. However, its physical interpretation is still under discussion (for reviews, see Refs.~\cite{Wiltshire:1995vk,Halliwell:1989myn,Kiefer:2008sw}). Additionally, there is an ambiguity in operator ordering during the quantization of the classical Hamiltonian \cite{Hawking:1985bk,Kontoleon:1998pw}.

To solve the Wheeler-DeWitt equation, a framework known as a minisuperspace model, which restricts the infinite degrees of freedom of spacetime to a finite number by imposing symmetries, is often used.
In particular, the system with a canonical scalar field in a homogeneous and isotropic Universe has been studied in detail \cite{Blyth:1975is,Esposito:1988aw,Garay:1990re,Gibbons:1989ru,Guzman:2005xt}. In this study, we analyze the Wheeler-DeWitt equation under the requirement $|\Psi|=1$. This condition describes the limit where the classical Hamilton-Jacobi equation is recovered from the Wheeler-DeWitt equation \cite{Vink:1990fm}. In the de Broglie-Bohm interpretation, which is characterized by the deterministic time evolution, the condition $|\Psi|=1$ can be understood as the condition where the quantum trajectories exactly coincide with classical ones. (See Ref.~\cite{Pinto-Neto:2013toa} for the review of quantum cosmology with de Broglie-Bohm interpretation). The theoretical importance of such solutions was emphasized in \cite{John:2014sea}. Assuming a perfect fluid satisfying an energy density $\rho\propto a^{-n}$ (where $n$ is a constant) as the matter component, they discovered analytical solutions ($n=2$) to the Wheeler-DeWitt equation that also hold classically. However, their results are restricted to a single dynamical degree of freedom (the scale factor $a$) and rely on a fixed operator ordering.

In this study, we derive the most general analytical solutions that satisfy the requirement $|\Psi|=1$ for a two-degree-of-freedom system $(a,\phi)$ in a flat, homogeneous, and isotropic spacetime without specifying the operator ordering. Consequently, we have shown that the scalar field potential $V(\phi)$, which is completely arbitrary in classical theory, is fully determined from the imaginary part of the Wheeler-DeWitt equation as a function of the operator ordering parameter. Depending on the value of the operator ordering parameter, the allowed potential classes include the exponential potential, the quadratic potential, and a cosine-type potential combined with negative cosmological constant. Although our analysis is motivated by the de Broglie-Bohm interpretation, this classification of allowed solutions and potentials under the condition $|\Psi|=1$ does not rely on any specific interpretation. 

Additionally, focusing on the cosine-type potential with a negative cosmological constant in the classified allowed potential classes, we explicitly derive the analytical solution for the scale factor and scalar field and discuss the cosmological implications.

 In Sec.~\ref{sec:Formalism}, we introduce the theoretical framework of quantum cosmology, and in Sec.~\ref{sec:Classification_of_the_Allowed_Potentials}, we classify allowed potentials under the condition $|\Psi|=1$. In Sec.~\ref{sec:Solutions_To_the_Cosine-type_Potential_with_a_Negative_Cosmological_Constant}, we derive analytical solutions for the case of cosine-type potential with a negative cosmological constant. Sec.~\ref{Conclusion} is devoted to the conclusion. Appendix \ref{sec:Derivation_of_The_General_Form_of_the_S} contains the rigorous proof for the general form of the phase of the wave function under the condition $|\Psi|=1$. 
 
 We choose the unit system where $8\pi G=1$ and $\hbar=1$ throughout this paper.

\section{Formalism}
\label{sec:Formalism}

For a homogeneous and isotropic Universe with a canonical scalar field $\phi$, the classical Lagrangian is given by 
\begin{align}
L(\phi,\dot{\phi},\alpha,\dot{\alpha})=e^{3\alpha}\left( \frac{1}{2}\dot{\phi}^2-V-3\dot{\alpha}^2 \right),
\end{align}
where $\alpha\equiv \ln a$ is the e-folding number and $V(\phi)$ is a scalar field potential. Since the canonical momenta are given by
\begin{align}
  &\pi_\alpha=\frac{\partial L}{\partial \dot{\alpha}}=-6\dot{\alpha}e^{3\alpha}\\
  &\pi_\phi =\frac{\partial L}{\partial \dot{\phi}}=e^{3\alpha}\dot{\phi},
\end{align}
the classical Hamiltonian becomes
\begin{align}
    H=e^{-3\alpha}\left( \frac{1}{2}\pi_\phi^2-\frac{1}{12}\pi_\alpha^2 \right)+e^{3\alpha}V.
\end{align}
Applying the canonical quantization,
\begin{align}
  \pi_\alpha\to -i\frac{\partial}{\partial\alpha},\quad \pi_\phi\to -i\frac{\partial}{\partial\phi},
\end{align}
we obtain the Wheeler-DeWitt equation $H\Psi=0$ as
\begin{align}
  -\frac{1}{12}\frac{\partial^2\Psi}{\partial\alpha^2}+\frac{1}{2}\frac{\partial^2\Psi}{\partial\phi^2}+q\frac{\partial\Psi}{\partial\alpha}-e^{6\alpha}V\Psi=0,
  \label{eq:Wheeler-DeWitt}
\end{align}
where $q$ is a real constant parameter for ambiguity in operator ordering. For instance, if we adopt the Laplace-Beltrami ordering \cite{Kiefer:2008sw}, $q=0$. In the following, we proceed with our discussion without specifying a particular value for $q$. Furthermore, expressing the wave function as $\Psi=|\Psi|e^{iS}$ with a real function $S$, we obtain
\begin{align}
    \frac{1}{12}(\partial_\alpha S)^2-\frac{1}{2}(\partial_\phi S)^2-e^{6\alpha}\left( V-\frac{1}{2e^{6\alpha}}\frac{\partial_\phi^2|\Psi|}{|\Psi|}+\frac{1}{12e^{6\alpha}}\frac{\partial_\alpha^2|\Psi|}{|\Psi|}-q\frac{\partial_\alpha|\Psi|}{e^{6\alpha}|\Psi|}\right)=0,
    \label{eq:real_part_of_Wheeler-DeWitt_equation}
\end{align}
as the real part of the Wheeler-DeWitt equation (Eq.~\eqref{eq:Wheeler-DeWitt}). Assuming $|\Psi|=1$, this equation exactly becomes the classical Hamilton-Jacobi equation:
\begin{align}
    \frac{1}{12}(\partial_\alpha S_{\mathrm{cl}})^2-\frac{1}{2}(\partial_\phi S_{\mathrm{cl}})^2-e^{6\alpha}V=0.
    \label{eq:classical_HJ_equation}
\end{align}
Here, $S_{\mathrm{cl}}$ is Hamilton's principal function, satisfying 
\begin{align}
    \pi_\phi=\frac{\partial S_{\mathrm{cl}}}{\partial\phi},\qquad \pi_\alpha=\frac{\partial S_{\mathrm{cl}}}{\partial \alpha}.
    \label{eq:HJ_momentum_relations}
\end{align}
Therefore, the condition $|\Psi|=1$ can be understood as the limit where the real part of the Wheeler-DeWitt equation (Eq.~\eqref{eq:real_part_of_Wheeler-DeWitt_equation}) reduces to the classical Hamilton-Jacobi equation (Eq.~\eqref{eq:classical_HJ_equation}). While the discussion so far does not rely on any specific interpretation of the wave function, the Bohmian interpretation assumes that Eq.~\eqref{eq:HJ_momentum_relations} are valid even in the quantum Hamilton-Jacobi equation (Eq.~\eqref{eq:real_part_of_Wheeler-DeWitt_equation}). In this framework, the deterministic time evolution is determined by the so-called guidance equations:
\begin{align}
    \pi_\phi=\frac{\partial S}{\partial\phi},\qquad \pi_\alpha=\frac{\partial S}{\partial \alpha}.
\end{align}
The primary advantage of this interpretation is that it does not face the problem of an external observer.

\section{Classification of the Allowed Potentials}
\label{sec:Classification_of_the_Allowed_Potentials}
In standard quantum cosmology, one typically assumes a specific potential $V(\phi)$ and solves for the wave function $\Psi$. In this section we adopt an inverse approach. Starting from the requirement $|\Psi|=1$, we derive the allowable potential $V(\phi)$. Under the condition $|\Psi|=1$, we obtain the following potential form:
\begin{align}
    V(\phi)=e^{-6\alpha} \left\{ \frac{1}{12}(\partial_\alpha S)^2-\frac{1}{2}(\partial_\phi S)^2 \right\}.
\end{align}
Since the potential does not depend on $\alpha$, the $\alpha$-dependence on the right-hand side must be exactly canceled. This means that the $\alpha$-dependence of $S$ should take the form $S\propto e^{3\alpha}$. Therefore, we can obtain following general form of $\Psi$ under the condition $|\Psi|=1$
\begin{align}
    \Psi&=e^{iS},\\
    &=e^{ie^{3\alpha}f(\phi)},
    \label{Psi}
\end{align}
where $f(\phi)$ is an arbitrary function of $\phi$. See Appendix~\ref{sec:Derivation_of_The_General_Form_of_the_S} for a more rigorous derivation of this form of $\Psi$. Substituting Eq.~\eqref{Psi} into 
the Wheeler-DeWitt equation Eq.~\eqref{eq:Wheeler-DeWitt}, we obtain
\begin{align}
    ie^{3\alpha}\left\{ -\frac{3}{4}f(\phi)+\frac{1}{2}f''(\phi)+3f(\phi)q \right\}\Psi+e^{6\alpha}\left\{ \frac{3}{4} f^2(\phi) -\frac{1}{2}\left( f'(\phi) \right)^2-V \right\}\Psi=0.
\end{align}
Here from the real and imaginary parts, we find
\begin{align}
    \frac{3}{4}f(\phi)^2-\frac{1}{2}\left( f'(\phi) \right)^2=V,\\
    f''(\phi)=6\left( \frac{1}{4}-q \right)f(\phi).
\end{align}
The imaginary part provides a constraint equation for the arbitrary function $f(\phi)$. The solutions for $f(\phi)$ are classified into three cases separated by the boundary at $q=1/4$. Table~\ref{tab:classification} summarizes the three classes of $f(\phi)$ and their corresponding potential.
\begin{table}[htbp]
  \centering
  \begin{tabular}{|l||c|c|}
    \hline
     & $f(\phi)$ & $V(\phi)$ \\
    \hline
    $q<1/4$ & $f(\phi)=Ae^{k\phi}+Be^{-k\phi}$ & $3q \left( A^2 e^{2k\phi}+B^2 e^{-2k\phi} \right) +6\left( \frac{1}{2}-q \right)AB$ \\
    \hline
    $q=1/4$ & $f(\phi)=A\phi+B$ & $\frac{3}{4}A^2\phi^2 +\frac{3}{2}AB\phi +\frac{3}{4}B^2 -\frac{1}{2}A^2$ \\
    \hline
    $q>1/4$ & $f(\phi)=A\sin (\omega \phi+B)$ & $A^2 \left\{ \left( \frac{3}{8}-\frac{\omega^2}{4} \right) - \left( \frac{3}{8}+\frac{\omega^2}{4} \right) \cos (2\omega \phi+2B) \right\}$ \\
    \hline
  \end{tabular}
  \caption{Classification for the solutions. Here, we defined $\omega\equiv\sqrt{6\left( q-\frac{1}{4} \right)}$ and $k\equiv\sqrt{6\left( \frac{1}{4}-q \right)}$.}
  \label{tab:classification}
\end{table}

The obtained classes of potentials encompass phenomenologically well-motivated potentials. For example, in the case of $q<1/4$, setting $B=0$ leads to the exponential potential
\begin{align}
    V=V_0 e^{-\lambda\phi},
\end{align}
where $\lambda=2\sqrt{6\left( \frac{1}{4}-q \right)}$. Such an exponential potential, which is widely investigated in the context of inflation and dark energy \cite{Halliwell:1986ja,Maki:2026qpf}, admits an asymptotic solution with an equation of state parameter $w=-1+\frac{\lambda^2}{3}$. To achieve accelerated cosmic expansion ($w<-1/3$) in this asymptotic state, the condition $\lambda<\sqrt{2}$ is required. Therefore, we find that $q>1/6$ is necessary to realize acceleration with this exponential potential. Furthermore, in the case of $q=1/4$, choosing $A=-\lambda$ ($\lambda$ is constant) and $B=0$ yields 
\begin{align}
    V(\phi)=\frac{3}{4}\lambda^2\phi^2-\frac{\lambda^2}{2},
\end{align}
which possesses an analytic solution given by \cite{Lin:2023sza}
\begin{align}
    \phi=-\lambda t+C.
\end{align}
Since the scalar field $\phi$ evolves at a constant rate, this model is referred to as uniform rate inflation \cite{Lin:2023xgs,Lin:2023ceu}. 

Moreover, for $q>1/4$, the potential takes the form of a cosine-type potential plus a negative cosmological constant,
\begin{align}
    V=\Lambda-V_0 \cos (2\omega \phi +2B),
\end{align}
where $\Lambda\equiv \frac{3}{2}A^2(\frac{1}{2}-q)$ and $V_0\equiv \frac{3}{2}A^2 q$.
Regardless of how the constants and $q$ are chosen, it cannot be cast into an axion-like potential (such as $1-\cos$). The minimum of the potential is always negative: $V_{\mathrm{min}}=\frac{3}{2}A^2\left( \frac{1}{2}-2q \right) <0$.

\section{Solutions To the Cosine-type Potential with a Negative Cosmological Constant}
\label{sec:Solutions_To_the_Cosine-type_Potential_with_a_Negative_Cosmological_Constant}
\begin{figure}[htbp]
  \centering
  \includegraphics[scale = 0.6]{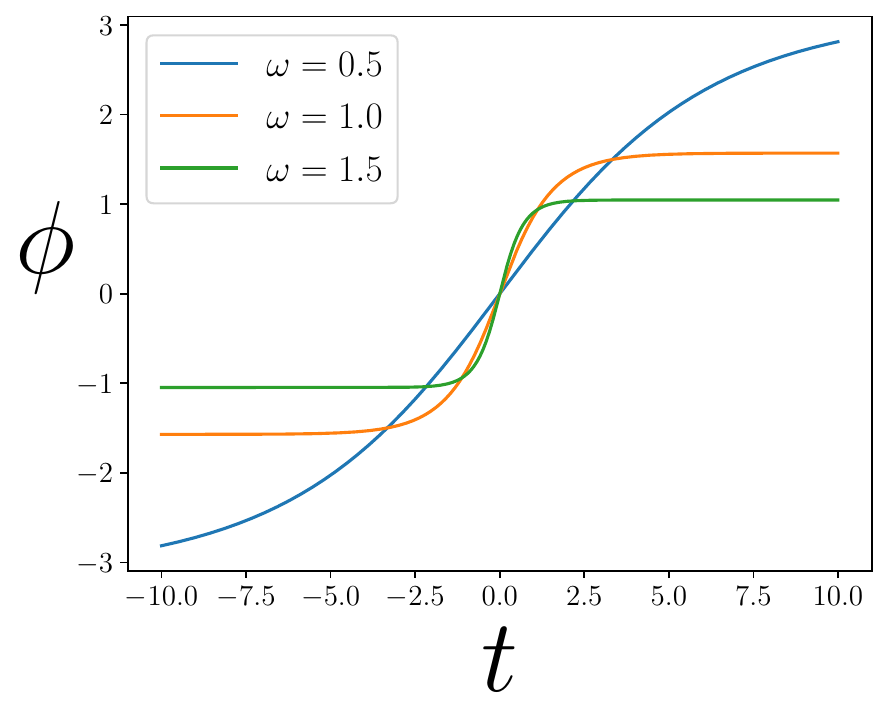}
  \caption{Evolution of $\phi$ for $\omega=0.5, 1.0, 1.5$, $A=1$ and $B=C=0$. In the limit of $t\to \pm \infty$, $\phi$ approaches $\pm \pi/2\omega$.}
  \label{fig:phi}
\end{figure}

In this section, we derive the analytical solutions for the case $q>1/4$. As shown in Table~\ref{tab:classification}, the function $f(\phi)$ and the corresponding scalar potential $V(\phi)$ are given by
\begin{align}
    f(\phi)&=A\sin (\omega \phi+B),\\
    V(\phi)&=A^2 \left\{ \left( \frac{3}{8}-\frac{\omega^2}{4} \right) - \left( \frac{3}{8}+\frac{\omega^2}{4} \right) \cos (2\omega \phi+2B) \right\},
\end{align}
where $\omega =\sqrt{6\left( q-\frac{1}{4} \right)}$, and $A$ and $B$ are arbitrary constants. Recall that the requirement $|\Psi|=1$ yields the phase form $S(\alpha,\phi)=e^{3\alpha}f(\phi)$ and the real part of the Wheeler-DeWitt equation exactly becomes classical Hamilton-Jacobi equation. Thus, the relations $\pi_\phi=\partial_\phi S$ and $\pi_\alpha=\partial_\alpha S$ can be understood either as the guidance equations in the Bohmian interpretation or as the classical relations in classical Hamilton-Jacobi theory. From the relation $\pi_\phi=\partial_\phi S$, we obtain
\begin{align}
    \dot{\phi}&=\frac{\partial_\phi S}{e^{3\alpha}},\\
  &=A\omega \cos (\omega \phi+B).
\end{align}
Integrating this equation yields
\begin{align}
    \phi=\frac{1}{\omega}\arcsin\left\{ \tanh(A\omega^2t+C) \right\}-\frac{B}{\omega},
\end{align}
where $C$ is an arbitrary constant of integration. Fig.~\ref{fig:phi} illustrates the evolution of $\phi$. 

\begin{figure}[htbp]
  \centering
  \includegraphics[scale = 0.6]{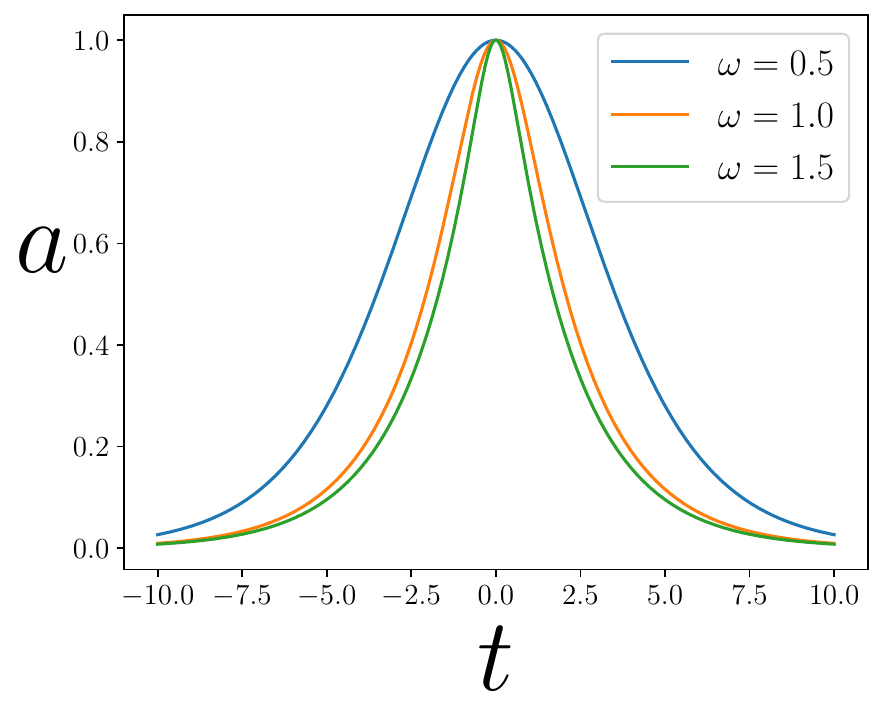}
  \caption{Evolution of the scale factor $a$ for $\omega=0.5,1.0,1.5$, $A=1$ and $C=0$. In the limit of $t\to \pm \infty$, $a$ approaches $0$.}
  \label{fig:scale_factor}
\end{figure}

In the limits $t\to \pm\infty$, the field $\phi$ approaches the asymptotic values $\phi=\pm\pi/2\omega$. This solution represents a scenario where the scalar field evolves from one potential maximum to the next over an infinite time. Next, we derive the analytical solution for the scale factor. From the relation $\dot{\alpha}=-\frac{1}{6}e^{-3\alpha}\partial_\alpha S=-f/2$, we obtain
\begin{align}
  \dot{\alpha}&=-\frac{A}{2}\sin(\omega\phi+B),\\
  &=-\frac{A}{2}\sin\left\{ \arcsin\left( \tanh(A\omega^2t+C) \right) \right\},\\
  &=-\frac{A}{2}\tanh(A\omega^2 t+C).
\end{align}
Integrating this yields
\begin{align}
  a&=\exp\left( -\frac{1}{2\omega^2}\ln \left\{ \cosh\left( A\omega^2 t+C \right) \right\} \right),\\
  &=\left\{ \cosh(A\omega^2t +C) \right\}^{-1/2\omega^2},
\end{align}
where we set $a=1$ at $t=-C/A\omega^2$. Fig.~\ref{fig:scale_factor} shows the evolution of the scale factor $a$. In the limit of $t\to \pm \infty$, this exact solution exhibits a de Sitter-like expansion and contraction $a\propto e^{\mp At/2}$. The scale factor decelerates ($\ddot{a}<0$) for $a>\left( 1+2\omega^2 \right)^{-1/4\omega^2}$, and accelerates ($\ddot{a}>0$) for $a<\left( 1+2\omega^2 \right)^{-1/4\omega^2}$. Because of this acceleration in the regime $a<\left( 1+2\omega^2 \right)^{-1/4\omega^2}$, the Universe avoids reaching the singularity ($a=0$) within a finite time. A similar time-symmetric solution, which also approaches a singularity over infinite time, was obtained for the $q=1/4$ case in \cite{Lin:2023sza}.

\section{Conclusion}
\label{Conclusion}
In this study, we have investigated the Wheeler-DeWitt equation with a scalar field in a flat, homogeneous and isotropic spacetime. We have shown that under the requirement $|\Psi|=1$ and the assumption that the scalar potential $V$ is independent of the scale factor, the general solution of the wave function $\Psi$ takes the form $\Psi=e^{ie^{3\alpha}f(\phi)}$, where $\alpha$ is the e-folding number $\alpha=\ln a$ and $f(\phi)$ is an arbitrary function of $\phi$. Furthermore, we demonstrated that the imaginary part of the Wheeler-DeWitt equation completely determines $f(\phi)$ and the potential $V(\phi)$ depending on the operator ordering parameter $q$. The allowed potentials under $|\Psi|=1$ are classified by the operator ordering parameter $q$. We have shown that the classified potentials admit simple forms, such as the exponential, quadratic with a negative cosmological constant, and cosine-type potential with a negative cosmological constant. Furthermore, focusing on the case of $q>1/4$, we have derived analytical solutions for the scale factor and the scalar field and shown that the cosmological evolution avoids a singularity within a finite time. It would be interesting to explore inflationary or dark energy models based on these results, as they admit analytical solutions for the evolution of both the scale factor and the scalar field, without relying on approximation schemes such as the slow-roll conditions. We leave this investigation for future work.

\begin{acknowledgments}
This work was in part supported by the National Science and Technology Council (NSTC) of Taiwan under grant number 114-2112-M-167-001 (CML) and JSPS KAKENHI Grants
No. JP24K07027 (KK).
\end{acknowledgments}

\appendix

\section{Derivation of The General Form of the S}\label{sec:Derivation_of_The_General_Form_of_the_S}

Here, we use the Lagrange-Charpit method for first-order partial differential equations to prove that the form $S=e^{3\alpha}f(\phi)$ is naturally obtained by requiring $|\Psi|=1$ and that the potential $V$ is independent of $\alpha$, i.e., $V=V(\phi)$. Defining $u(\alpha,\phi)$ via the relation
\begin{align}
    S(\alpha,\phi)=e^{3\alpha}u(\alpha,\phi),
\end{align}
and rewriting the real part of the Wheeler-DeWitt equation (Eq.~\eqref{eq:real_part_of_Wheeler-DeWitt_equation})
\begin{align}
    V(\phi)=e^{-6\alpha}\left\{ \frac{1}{12}(\partial_\alpha S)^2-\frac{1}{2}(\partial_\phi S)^2 \right\},
\end{align}
as a partial differential equation for $u$, we obtain
\begin{align}
  F(\alpha,\phi,u,p_\alpha,p_\phi)\equiv(3u+p_\alpha)^2-6p_\phi^2-12V(\phi)=0,
\end{align}
where $p_\alpha\equiv\frac{\partial u}{\partial \alpha}$, $p_\phi\equiv\frac{\partial u}{\partial \phi}$. 
According to the Lagrange-Charpit method for the first-order partial differential equation, the following characteristic equations hold:
\begin{align}
  \frac{d\alpha}{\partial_{p_\alpha} F}=\frac{d\phi}{\partial_{p_\phi} F}=-\frac{dp_\alpha}{\partial_\alpha F+p_\alpha\partial_u F}=-\frac{dp_\phi}{\partial_\phi F +p_\phi\partial_u F}=\frac{du}{p_\alpha\partial_{p_\alpha}  F +p_\phi\partial_{p_\phi} F}\equiv ds.
\end{align}
This yields
\begin{align}
  &\frac{d\alpha}{ds}=2(3u+p_\alpha),\\
  &\frac{d\phi}{ds}=-12p_\phi,\\
  &\frac{dp_\alpha}{ds}=-6p_\alpha(3u+p_\alpha),\\
  &\frac{dp_\phi}{ds}=12\partial_\phi V-6p_\phi(3u+p_\alpha),\\
  &\frac{du}{ds}=2p_\alpha(p_\alpha+3u)-12p_\phi^2.
\end{align}
From the first and third equations, we find 
\begin{align}
  \frac{dp_\alpha}{d\alpha}=-3p_\alpha.
\end{align}
Solving this gives
\begin{align}
  p_\alpha=Ce^{-3\alpha},
\end{align}
where $C$ is an arbitrary constant. Consequently, we obtain
\begin{align}
  u(\alpha,\phi)=-\frac{C}{3}e^{-3\alpha}+f(\phi).
\end{align}
Here, $f(\phi)$ is a function of $\phi$. Note that $f(\phi)$ is not a completely arbitrary function. Substituting this into the original equation $F=0$ yields
\begin{align}
  V(\phi)=\frac{3}{4}f(\phi)^2-\frac{1}{2}\left\{ f'(\phi) \right\}^2.
\end{align}
$f(\phi)$ must be a solution to this equation. The phase $S(\alpha,\phi)$ is given by
\begin{align}
  S(\alpha,\phi)&=e^{3\alpha}u(\alpha,\phi)\\
  &=-\frac{C}{3}+e^{3\alpha}f(\phi).
\end{align}
Since an overall constant shift in $S$ does not change the Wheeler-DeWitt equation, we can set the general form for the phase $S$ under the condition of $|\Psi|=1$ as
\begin{align}
  S(\alpha,\phi)=e^{3\alpha}f(\phi),
\end{align}
without loss of generality.

\bibliography{reference}

\end{document}